# A heuristic approach to author name disambiguation in bibliometrics databases for large-scale research assessments[1]

Cristiano Giuffrida[*]

*Vrije Universiteit, Amsterdam, the Netherlands*

ADDRESS: Dept of Computer Science. Vrije Universiteit, De Boelelaan 1081A, 1081 HV Amsterdam - THE NETHERLANDS. tel. and fax +31 205987506, c.giuffrida@few.vu.nl

Ciriaco Andrea D'Angelo

*Laboratory for Studies of Research and Technology Transfer at
University of Rome "Tor Vergata" – Italy*

ADDRESS: Dipartimento di Ingegneria dell'Impresa, Università degli Studi di Roma "Tor Vergata", Via del Politecnico 1, 00133 Roma - ITALY, tel. and fax +39 06 72597362, dangelo@disp.uniroma2.it

Giovanni Abramo

*National Research Council of Italy and Laboratory for Studies of Research and Technology Transfer at
University of Rome "Tor Vergata" – Italy*

ADDRESS: Dipartimento di Ingegneria dell'Impresa, Università degli Studi di Roma "Tor Vergata", Via del Politecnico 1, 00133 Roma - ITALY, tel. and fax +39 06 72597362, abramo@disp.uniroma2.it

**Abstract**
National exercises for the evaluation of research activity by universities are becoming regular practice in ever more countries. These exercises have mainly been conducted through the application of peer-review methods. Bibliometrics has not been able to offer a valid large-scale alternative because of almost overwhelming difficulties in identifying the true author of each publication. We will address this problem by presenting a heuristic approach to author name disambiguation in bibliometric datasets for large-scale research assessments. The application proposed concerns the Italian university system, consisting of 80 universities and a research staff of over 60,000 scientists. The key advantage of the proposed approach is the ease of implementation. The algorithms are of practical application and have considerably better scalability and expandability properties than state-of-the-art unsupervised approaches. Moreover, the performance in terms of precision and recall, which can be further improved, seems thoroughly adequate for the typical needs of large-scale bibliometric research assessments.

**Introduction**

The current era of the knowledge economy demands that the governments of advanced nations devote ever more attention to the improvement of scientific-technological infrastructure, in order to sustain the competitiveness of national industry. Publicly funded research organizations (PFROs), such as universities and research institutes, take on a determining role in this context. Thus it is not surprising that in an ever greater number of nations, these organizations are the object of national research evaluation exercises. The exercises serve towards four principal objectives, adopted in whole or in part by the governments concerned: i) stimulation of greater production efficiency; ii) selective funding allocations; iii) reduction of information asymmetry between supply and demand in the market for knowledge; and last but not least iv) demonstration that investment in research is effective and delivers public benefits.

Until recently, the conduct of these evaluation exercises has been founded on the so-called peer-review methodology, where research products submitted by institutions are evaluated by appointed experts.

Peer review presents two severe limits. The first is that it bases evaluation on a limited subset of research products from the PFROs. This seriously jeopardizes: i) robustness of the measurement system, as shown by sensitivity analysis of performance rankings to the share of product evaluated (Abramo et al., 2010); and ii) validity of the measure of performance, due to the inefficient selection of the best products by PFROs (Abramo et al., 2009). The second limitation, a consequence of the first, is that the evaluation cannot measure "productivity", the quintessential indicator of efficiency for any production system. Recent advances in bibliometric techniques can without doubt contribute to mitigating the limits of peer review. It is not by chance that the upcoming evaluation exercises by the governments of the UK (the Research Excellence Framework, REF, to be launched in 2012), Australia (Excellence in Research for Australia, ERA, just launched), and Italy (Quintennial Research Evaluation, VQR, now being launched) will make more or less exclusive use of bibliometrics, where applicable (hard sciences).

The leap that policy makers and practitioners have been expecting from bibliometrics for some time is that: i) it would contribute at a large-scale measurement of the productivity of individual scientists, and through aggregation, of research groups, departments, faculties, disciplines, and PFROs as a whole; and ii) the measurement could be done avoiding submission of output data on the part of the PFROs[2]. A non-invasive approach in fact, in addition to significantly reducing the direct costs of evaluation this leap would also eliminate the indirect costs to the PFROs. Unfortunately, the severe technical problems associated with the bibliometric databases currently available, such as Thomson Reuters Web of Science (WoS) or Elsevier Scopus, make the objective difficult to realize. In the records of the publications, the authors are often indicated only by last name and initial of the first name, and there is no deterministic link between the "authors list" and the "address list"[3]. For every occasion that two or more affiliations are indicated it is not possible to identify which one of these is the specific affiliation of each author. In addition, for large populations of scientists, cases of homonyms are very numerous, and their disambiguation within acceptable margins of error is exceptionally difficult.

---

[2] An earlier experience of this kind in Australia, the Composite Index, demonstrated the problems encountered with output submission. Audits conducted by KPMG found a high error rate in publication lists submitted by universities, especially at the outset of the application of this approach (34% in 1997). This error rate caused 97% of errors in the final scores and consequent funding allocations (Harman, 2000), although there has recently been a notable reduction in these rates.

[3] Bibliometric records without these limitations have only become available in recent years.

The impact of name ambiguity varies depending on the area of application. In information retrieval systems, name ambiguity increases the amount of noise in search results. As an example, querying Google with the string "Mario Bernardi" results in more than 70,000 web pages, with seven different individuals represented in just the first 10 results: a tourist guide, a Canadian orchestra director, a car dealer, a researcher from the Faculty of Science of the University of Verona, a script writer, and an actor. In bibliometrics, name ambiguity represents a considerable source of error and can affect the quality and validity of the results. A recent study (Aksnes, 2008) has shown that 14% of Norwegian authors share their name with one or more other individuals. Examining papers published between 1893 and 2006 in journals of the collection of Physical Review, Radicchi et al. (2009) estimated that 8% of the number of different physical authors had names that were homonyms. Prior research has suggested that the distribution of homonyms may be even more significant for other countries and languages (Cornell, 1982). This has led to a situation in the literature where bibliometric studies based on observation of data at the level of single scientists have been restricted to specific disciplinary sectors or single institutions, contexts where the quantity of data is very contained and it is possible to proceed through manual disambiguation of individual authorship[4].

Census on a name basis appears to be the fundamental step in overcoming many of the methodological limits of bibliometrics. Such an achievement permits measurement of research productivity at all levels desired: single individuals, but also departments, scientific sectors, disciplinary areas, faculties, or entire organizations. In fact, it is only through comparison of performance for productivity and impact in the same scientific sector (that of the author) that it is possible to compare individuals and, through successive aggregations, research units operating in different sectors, without encountering the distortions due to the differing scientific "fertility" of the disciplinary areas (Abramo et al., 2008a).

To the best of our knowledge, the only non-invasive bibliometrics-based attempt at comparative evaluation of university research productivity on a national scale is that of Abramo et al. (2008b), for the case of Italy. In this work, the authors presented a non-parametric bibliometric measurement of the production efficiency of Italian universities by disciplinary area, beginning from the scientific product of individual researchers. In the current work we will present the technology which underlies the method used and the manner in which it was developed, applied and validated. Currently, a number of Italian universities (e.g., the universities of Rome "Tor Vergata", Milan, Pavia, Cagliari Udine, and Luiss), the research institutes Fondazione Bruno Kessler and Edmund Mach, and the Sardinia administrative region have already used such research assessment system. The aim of this manuscript is not that of providing principled and lasting contributions to the problem of name disambiguation, rather tackling a new author disambiguation setting where author records from a bibliometric database are matched to the official records for research evaluation. The following section of this paper presents a summary of the state of the art for author name disambiguation in bibliometrics. In Section 3 we will illustrate the field of application and dataset used in the study, while Section 4 will present the general scheme and the algorithmic details of the heuristic approach proposed. Section 5 is dedicated to the presentation of results obtained in the validation tests for the approach, while the closing section provides a synthesis of the results and the final considerations of the authors.

---

[4] As an illustration, we report the comments by Van Raan (2008), a leading scholar in the field of bibliometrics, concerning his examination of a dataset of 18,000 WoS publication listings by all chemistry researchers in 10 Dutch universities: "This material is quite unique. To our knowledge, no such compilations of very accurately verified publication sets on a large scale are used for statistical analysis of the characteristics of the indicators at the research group level".

**Approaches to author disambiguation in bibliometrics**

Author name disambiguation is a special case of entity disambiguation where entities are not labeled with unique identifiers. The general problem has now been studied for years, with approaches tailored to both structured and unstructured data, proposed in several different research communities. This work is concerned with the problem of author name disambiguation in structured data, specifically focusing on bibliometric applications. In bibliometric databases, articles are stored in the form of records including the list of the original authors and other attributes, such as the title of the article and the journal. Each article represents an author instance for each author in the author list.

The problem of author name disambiguation is generally broken down into two separate subtasks. The first task is separating the instances of multiple authors that share the same name. For example, "Francesco Adamo" may refer to a researcher in Electronics from Bari Polytechnic or a full professor in Economics from the University of Eastern Piedmont. The second task is identifying instances of an author with different names. For example, "Maria Ippoliti", "Valeria Ippoliti", and "Maria Valeria Ippoliti" may all refer to a researcher in Political Science from University of Salerno. Both problems are equally relevant in bibliometric databases[5] and can have several causes, including identical names, name variations, pseudonyms or alternate spellings, name abbreviations, misspellings, typographical errors, and OCR-originated (optical character recognition) errors. Most approaches described in the literature deal with both tasks.

Methods to disambiguate author names are usually categorized as supervised methods and unsupervised methods. Supervised methods require manually labeled data to train the algorithm before disambiguating each author instance. The training set can be used to learn characteristics of each author or a generic similarity metric between instances of the same author. The latter approach has been recently explored by Torvik et al. (2005) to build a probabilistic model for estimating the probability that a pair of author instances refer to the same individual. They evaluate their approach on the Medline database using a comparison vector across eight different attributes: middle initial, suffix, journal name, language, co-authorship, title, affiliation, MeSH words.

The vast majority of other supervised approaches use training data for each author to disambiguate. This method usually yields better results as the disambiguation algorithm has specific information on each ambiguous author. Another key advantage is the ability to leverage well-established machine learning technologies like classification and clustering. The most common approach is to first build a binary classifier that is able to predict whether a pair of instances refers to the same author. Then, a clustering algorithm that employs the classifier as a distance metric is used to group all the co-referent instances together. One of the most relevant developments in this direction is presented by Han et al. (2004). They propose two models: a generative model (Naive Bayes probability model) and a discriminative model (Support Vector Machines, or SVM). In both cases they focus on three attributes: co-authorship, title, journal name. Experimental results show that SVM outperforms Naive Bayes in general, but the latter is able to better model unseen information, especially coauthoring patterns. They also show that co-authorship represents the most important attribute to disambiguate author instances. Not surprisingly, other supervised approaches have concentrated exclusively on co-authorship information to disambiguate author instances. An example is co-author inclusion (CAI) (Wooding et al., 2006), an algorithm to recursively expand a core set of papers of an author based on co-authors found in the original set. Multi-attribute supervised approaches represent a generalization of this approach.

---

[5] These problems are often amplified by the tendency for authors to be indicated only by last name and single initial of the first name.

An important shortcoming of standard supervised approaches is the so-called transitivity problem. This problem stems from the use of a binary classifier that analyzes pairs of instances without considering the global distribution of an author's product. In practice, the collection of papers of an author is not completely homogeneous and the distance between two papers of the same author may vary. For this reason, many existing algorithms do not usually handle correctly papers that are not part of the author's main work. To overcome this issue, some approaches extend standard machine learning techniques with global metrics on a set of records (Culotta et al., 2007).

The major drawback of supervised approaches is the need for a training set. This assumption is expensive in practice, and manual labeling of data can become impractical for large-scale bibliometric databases. In addition, maintaining the training set may be prohibitive when data change frequently. To address these issues, many unsupervised approaches have been proposed. These methods do not need manually labeled data for training the disambiguation algorithm.

Many unsupervised approaches formulate the author name disambiguation problem as a clustering task, where each cluster contains all the articles by the same author. In this case, the distance metric is not learned by a training set but it is given directly by the model employed. As before, both generative models (Han et al., 2005a; Soler, 2007; Song et al., 2007) and discriminative models (Han et al., 2005b) have been investigated in the literature. In another direction, other unsupervised approaches have used collaboration or citation graphs to disambiguate authors. For example, McRae-Spencer and Shadbolt (2006) suggest the use of self-citation graphs to iteratively tie together articles of the same author. This approach yields very high precision but low recall due to articles that are outside an author's main citation network.

As discussed earlier, disambiguating authorship of publications which do not fall within an author's main discipline of research is more difficult in general. Both supervised and unsupervised approaches that use local metrics to disambiguate authors are greatly affected by this issue. Other important shortcomings in existing approaches are poor scalability and expandability properties. Most algorithms cannot be used efficiently in large-scale bibliometric databases and cannot handle frequent changes to the database properly. Some attempts to solve scalability issues use a two-step blocking framework to reduce clustering complexity (On et al., 2005; Huang et al., 2006). Online learning has also been proposed to improve expandability properties (Huang et al., 2006). These approaches, however, all yield an inherent implementation and maintenance complexity that may limit their practical application.

To address the challenges discussed above, we explore a different approach to author name disambiguation in structured data. Our approach stems from the intuition that the integration of heterogeneous data sources is natural and practical in bibliometric applications. Consider an analysis to generate the list of the top researchers and research institutes in a particular country. Data integration is necessary to build the analysis upon official lists and produce reliable and comparable results. The key idea behind our approach is to leverage data integration with external information sources to obtain more data on the authors and ease the task of author name disambiguation. Existing algorithms are largely vulnerable to lack of information in the bibliometric database. For example, an author publishing in different scientific areas may have articles with completely different titles, keywords and journals. By leveraging only internal information, his/her papers will likely end up in different clusters. The transitivity problem and the difficulties to identify articles not part of the author's main product are also often a consequence of lack of information.

The idea of exploiting data integration for disambiguating instances of an author is not entirely new. For example, Li et al. (2005) present a name disambiguation algorithm using integration of semi-structured and unstructured information. Other studies have proposed the integration of data from the web to ease the task of author name disambiguation. Tan et al. (2006) suggests querying a web search engine with the title of each article and integrating the list of the first $r$ relevant URLs as a new attribute for each record. They show how the use of the resulting enriched attribute vector in a standard clustering algorithm greatly enhances the performance of the disambiguation process. In another direction, Kanani et al. (2007) use a web search engine query formed by conjoining the title of two articles and leverage the results to acquire information on the relationship of the two articles. They also present a model for resource-bounded information gathering from the web to keep the algorithm scalable.

In our approach, structured data is integrated with the bibliometric database in the most natural way. We use a reference external source of information that provides data on the institutional affiliation and research area of each Italian academic scientist. In most bibliometric applications, this assumption holds already in practice, as discussed earlier. In addition, we specifically tailor our design to bibliometric applications. We observe that false negatives (articles not assigned to their real authors) are less tolerable than false positives (articles erroneously assigned to the wrong identities). As a result, our algorithm is specifically optimized for assuring a high level of recall, but without losing in precision[6]. This is in contrast to the majority of existing approaches described in the literature that favor precision over recall.

Our approach follows a three-step process: database integration, mapping generation, and filtering. First, information from the "external" database is integrated into the bibliometric database. As a result, a reference list of author identities and their attributes is added to the original database. Second, a mapping algorithm links each author of an article to all the possible author identities from the reference list. Finally, different data-driven heuristics are used to filter out as many false positives as possible. The result of the last step is a robust mapping between author instances and author identities with a minimum number of false positives and a negligible number of false

---

[6] As shown below, the proposed algorithm guarantees values of precision and recall that are fully comparable.

negatives.

Other scholars have used a multi-stage process (Iversen, Gulbrandsen & Klitkou, 2007) to match authors in patent databases with an external database of inventors. Albeit similar in spirit to our methodology, their approach consists in a multi-step refinement, using only manual inspection with no explicit algorithm to automatically disambiguate authors. Using manual inspection to disambiguate author names is very effective for small populations of scientists, as other similar studies have also demonstrated (Meyer, 2003; Balconi, Breschi, & Lissoni, 2004). In large-scale applications, however, it is necessary to automate the disambiguation process as much as possible, to keep the approach feasible and easy to maintain over time, as more and more data becomes available.

The key advantage of our approach is the ease of implementation. As a result, our algorithms are of practical application and have considerably better scalability and expandability properties than state-of-the-art unsupervised approaches. Our approach provides appealing expandability properties, since it requires only minimal manual information that is essentially stable over time to process an evolving bibliometric dataset. As far as scalability properties are concerned, our technique has proven very effective in large-scale research assessments, since the disambiguation algorithm is entirely linear in the number of tuples processed. In addition, our algorithms consider global metrics and are therefore not exposed to the transitivity problem discussed earlier. Furthermore, the use of an external data source compensates internal noise and makes the approach less vulnerable to lack of information. Our approach is robust and applicable to different datasets and is resilient to the differences between an author's main work and his minor product. Finally, our method provides a viable way to efficiently integrate external data into existing bibliometric databases and provide a cleaner and richer dataset for bibliometric applications.

**Field of application and sources**

The application and validation of the heuristic approach, presented in detail in Sections 4 and 5, is carried out on the Italian university system. This section presents the databases used for the purpose.

**The bibliometric database**

The bibliometric database used for application of the disambiguation algorithm proposed by the authors consists of the 2001-2007 publications listed in the Thomson Reuters Italian National Citation Report (I-NCR). This assemblage is derived from the WoS database by considering bibliometric records with at least one address matching the term "Italy". It includes 333,743 publications, of which 271,296 are articles or reviews[7]. Their division by WoS macro-categories[8] is presented in Table 1.

---

[7] The dataset excludes other types of publications (editorials, letters, meeting abstracts, etc.) since they can not confidently be associated with a true research product, and would thus be a source of noise for any possible subsequent bibliometric evaluation.

[8] The WoS classifies the indexed publications by subject categories. The subject categories are grouped into macro-categories.

[Table 1]

In addition to a series of bibliometric data that are not relevant to the present ends, the I-NCR records present the following information for each publication:
- title
- abstract
- keywords
- type (article or review)
- journal
- volume, issue, pages
- year
- author list
- address list
- subject category.

The first six fields are de facto not relevant to the application of the proposed algorithm. The other four, in contrast, are used in the phases of database integration and mapping generation when merging data from the initial bibliometric database and the external database that we illustrate in the next section.

**The external database**

The Italian university system imposes a classification of research staff into 14 University Disciplinary Areas (UDAs) that, in turn, consist of 370 Scientific Disciplinary Sectors (SDSs)[9]. Each researcher is assigned to one and only one of these sectors. In practice, the hard sciences, representing 9 of the 14 UDAs, contain almost 64% of the total Italian university research staff, as seen in Table 2. Social sciences and Arts & humanities each account for 18%. The two classifications, namely the WoS categories and the UDAs, are clearly different. Not less, comparison of the data in Table 2 and Table 1 shows that for the Social sciences and Arts & humanities (together employing 36% of Italian researchers) the publications in the corresponding WoS categories represent only slightly more than 3% of the total, indicating the limited pertinence of the bibliometric approach in these areas. In any case, we will not exclude these areas from the application of the proposed algorithm[10].

[Table 2]

The external database considered is the database of research staff at all Italian universities for each year, beginning from 2000, as managed and updated by the CINECA university consortium on behalf of the Ministry of Education, Universities and Research[11]. For every researcher, in addition to the name, the database provides the following information:
- SDS,

---

[9] For a complete listing see http://www.miur.it/atti/2000/alladm001004_01.htm
[10] Our algorithm can be applied to any data source having properly structured records (e.g., book, book chapter, poetry) with at least three fundamental pieces of information: i) the author list; ii) the affiliation; iii) the subject category.
[11] See http://cercauniversita.cineca.it/php5/docenti/cerca.php.

- university,
- department or faculty within the university,
- official academic rank (assistant, associate, full professor).

As can better be seen in Section 4, only the first two fields are relevant to the application of the proposed algorithm. All the fields are categorized on an annual basis and currently report data as of December 31 of each year. This level of granularity is sufficient for the vast majority of bibliometric applications, given that the number of researchers that change affiliation or academic rank more than once a year is practically negligible[12].

Overall, the field of observation is composed of over 80 Italian universities. As of 31/12/2006, the total research personnel of these universities amounted to 62,559 individuals, among assistant, associate and full professor roles. These are the only individuals considered, being those who hold permanent positions responsible for research[13]. Other non-permanent figures, such as postdoctoral researchers, are being integrated into the database, as part of our ongoing work.

**The proposed approach**

The proposed approach is articulated in three distinct phases: database integration, mapping generation and filtering. The subdivision in phases and sub-phases permits a modular design for the algorithm, in which each step can be optimized or replaced under strategies suited to needs. Such a design guarantees high levels of control and configurability for the algorithm and provides a general model that can be readapted to a vast range of diverse settings. We will now discuss the various phases of the algorithm in detail, highlighting the design choices taken to reach the fundamental objective of maximizing recall while at the same time obtaining a high level of precision.

**Database integration**

The objective of this first phase is the integration of the bibliometric database with the external database. The ultimate goal is that of compensating for noise and lack of information in the bibliometric database, which contains references to author names with undefined identities. Ideally, an external source of information should consist of a subset with maximum coverage of the starting set of authors. In our case, the bibliometric database contains the articles by those authors who indicated some form of Italian affiliation, while the external database identifies all researchers on staff at Italian universities over the same period. The external database thus excludes university researchers without a formal position (assistant, associate, full professor), scientists of non-university research organizations in Italy, and foreign-based coauthors of publications listed in the bibliometric source. Below, we will describe the choices made to confront and resolve this problem. We note that, in practice, the extent of difficulty in the phase of integrating bibliometric data with external sources of information can be expected to depend on the compatibility between the formats of the two databases to be integrated. The applicability of our methodology to citation analysis setting different from Italy depends on the availability of an authority external database. In countries with public university systems this should be easily available, as shown by the various national research assessments which rely on them.

---

[12] We can assume that in other countries too, the event of researchers changing affiliation more than once a year, is negligible.

[13] The dataset thus excludes fellowship holders, research doctorates, students in post-doc and specialization stages and all other non-faculty figures.

**Mapping generation**

The objective of this phase is to generate a mapping of the authors present in the starting bibliometric database and the identities of the university researchers indexed in the external database. The output of the algorithm in this phase is a series of pairs (author, identity) that should represent a superset of the set of real pairs. The superset can contain, for every author extracted from the starting database, different pairs that compose the mapping with the multiple identity possibilities indexed in the external database. In contrast, the real mapping set that we are seeking must contain no more than one pair for each author.

To generate the mapping superset the algorithm employs strategies of aggressive matching between author and candidate identities. The objective is to map every author under all the possible candidate identities, of which at most one represents the real identity. The only authors excluded from the mapping are those that do not possess a candidate identity in the external database. Such authors are excluded from the subsequent steps of the disambiguation algorithm. The extent of occurrence of these "orphan" authors is conditional on the choice of the external information source. Their occurrence is minimal, for example, if the external database provides a subset of the starting set of authors that is a good approximation of the original set. In such cases, the exclusion of orphan authors should not only be marginal, but could also be convenient for bibliometric purposes. The dataset chosen for the current work, for example, automatically excludes all authors without official faculty roles in Italian universities. Since the final objective is actually to have a robust dataset for reliable bibliometric analysis of the very university system, this operation contributes to eliminating some of the noise that would be present in the analysis.

To implement an aggressive matching, the algorithm explores and maps all possible candidate identities for a given author. The match takes place between the name of the author and that of the corresponding identity. The form of the two names can be different, and for this reason the algorithm explores all possible criteria for matching. In the dataset chosen for the present work, the name of the author is in the form "SURNAME INITIALS" (e.g. ROSSI Giovanni Maria is listed as ROSSI GM), while the name of the identify is listed in full form. Unfortunately, the use of initials to represent the first name of the author causes considerable amplification of the problem of homonyms, and the complexity of the resulting disambiguation process. Furthermore, manual checks have demonstrated that, with cases of multiple first names, some of these may not be indicated in one of the two databases or can be indicated in a different order. To avoid exclusion of any possible mapping, the proposed algorithm assigns a relationship every time that there is a match between last names and when at least one initial of the first name of the author matches one of the first names of the identity. Our algorithm can be tuned to accept only exact matches between last names or establish a threshold on the minimum edit distance allowed between two matching strings. The latter scenario is important when spelling mistakes and OCR errors are to be expected. A higher threshold, however, will naturally generate a larger number of false positives.

In our experience, an approach that considers only almost exact matches between last names is effective in most cases. A second complication is presented by the possible presence of compound last names. This is the case, for example, of LEVIALDI GHIRON Nathan. This scientist, as author in the bibliometric database, can be listed in five different forms:
- LEVIALDI GHIRON N
- LEVIALDI N
- GHIRON N
- LEVIALDI GN
- GHIRON NL

The first three cases refer to the respective possibilities that the person opted to sign the publication using both or only one of his last names. The last two cases refer to the chance (not remote) that during the phase of indexing the bibliometric record, one of his last names was mistaken for a first name.

A third complication can be identified in the form in which an identity is defined in the external database. In typical bibliometric applications, such as with the dataset employed in the present work, the identities of each author are defined on an annual basis, without any correlation among identities that may pertain to different years. It is generally very complicated to trace the identity of a single author through the years in an automated manner, and without employing additional information. A large part of the complexity is due to changes in affiliations and to turnovers of term and tenured staff through the years, especially in extensive populations, such as the one considered in the proposed case. To address this challenge without abandoning the maximization of recall, the algorithm produces mappings between authors and identities based on the name and on an annual basis, with the final objective of disambiguating the authors and reconstructing the identity exclusively on a year by year basis. We point out, however, that this is not a shortcoming of our approach but rather a limitation of the external database employed. In ongoing work, we are evaluating other external data sources, from which it is possible to access or reconstruct traceability information about the authors.

**Filtering**

In the preceding phase the algorithm produced a set of mappings containing, in addition to the correct pairs (author, identity), a high number of false positives that, in the phase of filtering, the algorithm must eliminate. The false positives are all the possible cases of homonyms as produced from the phase of mapping generation. The cases of homonyms that it is possible to identify depend on the data available in the databases considered. In the dataset chosen for the present work, the possible cases of homonyms can be classified in the following categories.
- External homonyms: authors without a faculty role, or affiliated with some other type of research organization, with a homonym identity in a faculty role;
- Inter-address homonyms: authors in a university faculty role, having a homonym identity at another university;
- Intra-address homonyms: authors in a faculty role, with a homonym identity in the same university but belonging to a different SDS;
- Perfect homonym: authors in a faculty role having a homonym identity belonging to the same university and in the same SDS.

Every case of homonyms for a particular author produces a false positive, namely an excess pair of the form (author, identity) in the mapping set. The only exception is the case of perfect homonyms, where the occurrence is completely masked by the lack of information in the external database. Such cases of homonyms are not further disambiguated and the two identities become amalgamated as one, subsequent to the mapping generation phase. The occurrence of such cases is application-specific and their number depends on the quality of information in the external database. In real-life scenarios, experience shows a negligible incidence of such cases in bibliometric applications. In the chosen dataset, the number of cases in question is certainly negligible, with perfect homonyms equal to 0.043% of the total researches in the Italian university system.

All the pairs of the form (author, identity) pertaining to a given author form a cluster. All the clusters of cardinality one represent instances of authors classified as completely free of cases of homonyms or else as cases of external homonyms not yet resolved. In contrast, clusters of cardinality zero represent cases of orphan authors with no identity[14]. The elimination of the false positives generated in the mapping generation phase is the result of a step-by-step process, gradually filtering out undesired pairs. The filters employed follow data-driven heuristics and depend on the particular application under examination. In the proposed algorithm the filtering process is representative of the most typical bibliometric applications, but can easily be adapted to vast number of settings.

**The address filter**

The address filter eliminates all the (author, identity) pairs in which the author's affiliation (extracted from the "address" field of the bibliometric record) is incompatible with the identity's affiliation (the university identified for the researcher as listed in the external database). The effectiveness of the filter depends on the criteria employed for matching between the two affiliations, which are typically indicated in much different formats. The proposed algorithm employs rule-based criteria for matching based on a controlled vocabulary, able to immediately eliminate the greater part of external and inter-address homonyms, generating a negligible number of false negatives.

The method employed ensures high performance of the filter, at the cost of manual compilation and maintenance of the controlled vocabulary[15]. While the effort for the initial compilation is substantial, this approach still yields good expandability properties for typical bibliometric applications, in which data updated year by year present a variability which is high for articles and authors while negligible for affiliations[16].

Once the address filter is applied, the algorithm identifies all the clusters with non-zero cardinality and archives them for application of the subsequent filters.

---

[14] "Tuples" and clusters of this type could have resulted from the mapping generation phase or be directly produced by one of the subsequent filtering steps.

[15] Composed of over 30,000 rules that match over 89% of "Italy" addresses in the 2001-2007 I-NCR. Matches are not conducted for private clinics and enterprises, consortiums, foundations and for a minimal component (less than 1%) of public research organizations, for which disambiguation is not possible.

[16] A possible alternative that could reduce manual effort but maintain an acceptable level of performance would be to use supervised machine learning approaches. The evaluation of such techniques is part of our ongoing work.

**The WoS-SDS filter**

From all the (author, identity) pairs remaining after the previous filter, the WOS-SDS filter eliminates all those in which the WoS subject category of the article[17] published by the author is not compatible with the SDS for the identity. The idea is that an author that would publish an article in a certain subject category cannot possibly be associated with an identity that works in a completely different SDS. Again in this case, the effectiveness of the filter depends on the criteria for matching the two classifications. The proposed algorithm carries out the matching in deterministic fashion, on the basis of a purpose-prepared WoS-SDS mapping set. The WoS-SDS mapping is generated employing a set of manually compiled rules, integrated with additional data derived from manually verified pairs (of clusters of cardinality one). We remark that the effort involved in manual generation, while not inconsequential, must only be taken once, since the mapping remains substantially stable over time.

The filter is purposely conceived to capture and remove obvious cases of homonyms revealed by evident incompatibility of the disciplinary categories. The WoS-SDS mapping is intentionally constructed at a coarse level of granularity so as to minimize the production of false negatives.

Subsequently, more aggressive criteria for filtering are applied to the remaining clusters of cardinality greater than one. Such clusters represent cases of authors mapped with multiple identities that have survived the preceding filters. These obviously contain at least one false positive, which subsequent filters are designed to eliminate. The clusters of cardinality one, at the end of this step of the algorithm, are considered disambiguated and are not subjected to further application of filters. At this point in the process, the probability that these clusters contain external or inter-address homonyms can be considered negligible.

**The shared SDS filter**

The shared SDS filter eliminates all the (author, identity) pairs belonging to any cluster that contains a further pair where the identity has a shared SDS. A shared SDS is an SDS in common with another pair referring to the same article and already accepted by the algorithm as being correct. The idea is that the likelihood that two identities originating from two pairs associated with authors of the same article could belong to the same SDS depends on the possibility that the article is the result of collaboration between authors of the same SDS. Since collaborations within the same SDS are quite common, the filter can potentially eliminate a significant number of false positives.

The filter is expressly conceived to ensure a negligible production of false negatives. At the outset, to minimize the probability of false negatives, it is necessary that the SDS set have a sufficiently high cardinality: in the dataset used for the present study the SDS set has a cardinality of 370 and a distribution that is not notably polarized in specific sectors. Further, to minimize the level of introduced noise, the algorithm analyses shared SDSs only for pairs that have already been accepted as correct.

---

[17] Each article inherits all the WoS subject categories assigned to the journal. For articles published in Science, Nature e PNAS the editorial board of Thomson Reuters assigns specific categories by manually inspecting the content of the article. Articles published in other journals classified as "Multidisciplinary" in the WoS database account, in terms of Italian research products, for less than 0.3%, which is so small that its effect is negligible.

**The maximum correspondence filter**

The maximum correspondence filter is used to process all the remaining clusters of cardinality greater than one and thus address all the remaining cases of unresolved homonyms. For this purpose, for every cluster, the filter selects a single pair to retain and eliminates all others. The pair within each cluster that survives the filtering process is the one for which the SDS has maximum "correspondence" to the subject category of the article to which the cluster refers. The correspondence of an SDS to a particular subject category is defined (on the basis of a seed set) as the number of identities belonging to that SDS that result as authors of articles falling in the subject category. The algorithm uses a seed set constructed in automatic fashion, based on the authors of all the pairs already accepted as correct by the algorithm. In other words, the filter is conceived in such a manner as to analyze each cluster and select the correct pair based on an educated guess derived from the distribution of the previously disambiguated data. Due to this strategy, the filter is less accurate than the preceding filters and potentially susceptible to production of false negatives. However, such an approach is necessary in cases where the intention is to guarantee the total disambiguation of authors in the starting bibliometric database. In any case, the expected number of remaining non-unitary clusters is sufficiently small, following the application of the preceding filters, and the production of false negatives generally has an insignificant impact. The alternative is that of not employing the filter, at the cost of having a larger number of false positives in the final set.

In the final set, all the pairs are accepted by the algorithm as disambiguated and correct. However, the final set can still contain a certain number of false positives. Every remaining cluster with non-unitary cardinality, for example, includes at least one false positive. In addition, false negatives can produce orphan clusters lacking the original correct pair.

Figure 1 depicts the multi-stage process followed by the proposed algorithm. The following section, in turn, presents an illustrative example to demonstrate a step-by-step application of our methodology.

[Figure 1]

**Application**

Let us consider the following publication:
BOSCHERINI F, DADDATO S, GROPPO E, LAMBERTI C, LUCHES P, PRESTIPINO C, VALERI S, 2004. X-ray absorption study at the Mg and OK edges of ultrathin MgO epilayers on Ag(001). *Physical Review B*, (69)4, 045412, 1-9.

Below, we illustrate the step-by-step mechanism through which the algorithm, beginning from the seven authors of the publication, generates the respective identities and filters out the homonyms.

<u>Mapping generation</u>
Integration of information from the bibliometric record with the data from the external database produces four clusters with the following eight pairs:

- (BOSCHERINI F, BOSCHERINI Federico: FIS/01[18] - University of Bologna)
- (DADDATO S - D'ADDATO Sergio | MED/09 - University of Bologna)
- (DADDATO S - D'ADDATO Sergio | FIS/01 - University of Modena-Reggio Emilia)
- (LAMBERTI C - LAMBERTI Claudio | ING-INF/06 - University of Bologna)
- (LAMBERTI C - LAMBERTI Carlo | CHIM/02 - University of Turin)
- (LAMBERTI C - LAMBERTI Maria Carla | SECS-P/12 - University of Turin)
- (VALERI S - VALERI Sergio | FIS/01 - University of Modena-Reggio Emilia)
- (VALERI S - VALERI Stefano | L-ART/02 - University of Rome "La Sapienza")

Only the first pair presents cardinality of one. All the others present at least one false positive due to the effect of obvious homonyms. Note that the strategy of aggressive matching between authors and candidate identities has generated the pair (Lamberti C - Lamberti Maria Carla) in which the identity shows two first names, in evident contrast to the single first name of the author.

The addresses filter

The address list indicated in the bibliometric record is as follows:
- Univ Modena & Reggio Emilia, INFM, Natl Ctr Nanostruct & Biosyst S3, Modena, I-41100, Italy
- Univ Modena & Reggio Emilia, Dipartimento Fis, Modena, I-41100, Italy
- Univ Turin, INFM, UdR Torino, Turin, I-10125, Italy
- Univ Turin, Dipartimento Chim Inorgan Fis & Mat, Turin, I-10125, Italy
- Univ Bologna, Dipartimento Fis, Bologna, I-40127, Italy
- Univ Bologna, INFM, Bologna, I-40127, Italy

The authors, in addition to their university departments, have also indicated an affiliation with a research unit of the Italian National Institute of Condensed Matter (INFM). In any case, there are three distinct university addresses: the universities of Bologna, Torino and Modena-Reggio Emilia. Therefore, from the eight identities generated in the previous step, the address filter eliminates the one for VALERI Stefano, affiliated with the University of Rome "La Sapienza". The remaining seven are:
- BOSCHERINI Federico | FIS/01 - University of Bologna
- D'ADDATO Sergio | MED/09 - University of Bologna
- D'ADDATO Sergio | FIS/01 - University of Modena-Reggio Emilia
- LAMBERTI Carlo | CHIM/02 - University of Turin
- LAMBERTI Maria Carla | SECS-P/12 - University of Turin
- LAMBERTI Claudio | ING-INF/06 - University of Bologna
- VALERI Sergio | FIS/01 - University of Modena-Reggio Emilia.

These survive the filter and pass to the next step.

The WoS-SDS filter

The subject category associated with the article is "Physics, condensed matter". From the candidate identities, the filter only eliminates that of "LAMBERTI Maria Carla".

---

[18] This code represents the SDS the identity belongs to. In this particular case FIS/01 stands for "Experimental physics".

This researcher belongs to the SDS SECS-P/12 (Economic History), clearly incompatible with the subject category of the article. Of the remaining six identities, two are clusters with cardinality of one:
BOSCHERINI Federico | FIS/01 - University of Bologna
VALERI Sergio | FIS/01 - University of Modena-Reggio Emilia
These are accepted as disambiguated. The remaining identities, listed below, still include two false positives due to homonyms:
- D'ADDATO Sergio | MED/09 - University of Bologna
- D'ADDATO Sergio | FIS/01 - University of Modena-Reggio Emilia
- LAMBERTI Claudio | ING-INF/06 - University of Bologna
- LAMBERTI Carlo | CHIM/02 - University of Turin

The shared SDS filter
This filter acts on the pairs of the two clusters still to be disambiguated. The filter succeeds in resolving the ambiguity for author DADDATO S, since of the two candidate identities, that of D'ADDATO Sergio from the University of Modena-Reggio Emilia presents an SDS (FIS/01) shared with two other already disambiguated authors (BOSCHERINI F and VALERI S). The true identity of LAMBERTI C still remains to be disambiguated.

The maximum correspondence filter
This last filter must eliminate the false positive in the cluster for LAMBERTI C. The two candidate identities are:
- LAMBERTI Claudio | ING-INF/06 - University of Bologna
- LAMBERTI Carlo | CHIM/02 - University of Turin

The filter eliminates the first, since the correspondence of the relative SDS (ING-INF/06 - Electronic bioengineering and computer science) to the subject category of the publication (Physics, condensed matter) is clearly less than that of the other SDS (CHIM/02 - Physical chemistry).
The final list of disambiguated authors consists of four identities:
- BOSCHERINI Federico | FIS/01 – University of Bologna
- D'ADDATO Sergio | FIS/01 - University of Modena-Reggio Emilia
- VALERI Sergio | FIS/01 – University of Modena-Reggio Emilia
- LAMBERTI Carlo | CHIM/02 – University of Torino

In this specific case, the remaining three authors for the publication (GROPPO E, LUCHES P, PRESTIPINO C) were researchers without a formal faculty role in the university system, even though they were affiliated with the same organizations as the four disambiguated authors. Note that in the example presented, the algorithm produces a perfect round of disambiguation: no false positives and no false negatives are generated by the procedure.

Reviewing the application of the algorithm to the bibliometric dataset under examination, the most aggressive filter is definitely the first, dealing with the author addresses (Table 3). With respect to the results from the mapping generation phase, this filter eliminates almost two thirds of the identities (65.6%) and 17.9% of the publications. Except for possible false negatives, this concerns publications for which the authors are exclusively and all external or inter-address homonyms.

The WoS-SDS filter is not particularly selective but still reduces the identities for the paper by 7.2%.

The last two filters, as already stated, act only on the clusters with cardinality greater than one survived from the preceding filters, respectively removing 2% and 1.5% of the total of identities.

[Table 3]

**Measurement of precision and recall**

The performance of the proposed algorithm was measured using two different methods. The first is on the basis of a sample, involving extraction of a random set of publications and then manual testing for the presence of potential false positives and false negatives generated by the algorithm during the phase of disambiguating identities associated with the authors. The second method, concerning an entire segment of the population, involves a comparison with the records present in the official database of the University of Milan, containing all publications produced by the researchers of the university.

**Test on a sample basis**

Table 4 shows the parameters applied for extraction of the sample to be submitted to testing.

[Table 4]

Given these parameters[19], applying the sampling formula indicated in (1), the quantification of errors generated by the algorithm will require a number (n) of observations equal to 636.

$$n = \frac{N \cdot Z^2 \cdot p \cdot (1-p)}{(N-1) \cdot e^2 + Z^2 \cdot p \cdot (1-p)} \qquad (1)$$

We proceed to the random extraction of the publications necessary to obtain this number of observations[20]. The publications needed were 372. The test was carried out manually by searching the various sources available[21] for all the information necessary to identify all and only the identities that can be correctly associated with the authors of these publications.

It is thus possible to identify 28 false positives and 40 false negatives, corresponding to values of precision and recall of 95.6% and 93.8% and an f-measure of 94.7%.

The data presented here seem counter to our indication, in Section 2, that false negatives are less tolerable than false positives, and therefore that our algorithm would be specifically optimized to assure a high level of recall, but without losing precision. Nonetheless, the most frequently occurring cause of false negatives is the incorrect indication, on the part of the authors, of their true organizational affiliation (16 cases, equal to 40% of the total)[22]. A typical situation for the field of medicine concerns authors who, although holding a faculty role with a university department, indicate an affiliation with a clinic or hospital with which they collaborate. However, in 30% of the cases the error is instead generated by the removal of author-identity pairs caused by the WoS-SDS filter, which are cases in which the authors have published articles in subject categories extraneous to their official recognized SDS. A further 10% of the errors are caused by the address filter, particularly by errors present in the controlled vocabulary used to match addresses in the bibliometric database with official affiliations present in the external database.

---

[19] For population heterogeneity a "p" value of 12% is chosen, which is decidedly cautious, as will be seen, when compared to the real occurrence of errors from the algorithm.

[20] For each publication, there will be as many observations as there are correct (author, identity) tuples (true positives).

[21] In addition to drawing on the data present in the external database, the verification involved tracing the full name and exact affiliation of each author as indicated in the title area of the publication. In the remaining potentially doubtful cases (use of pseudonyms, etc.), additional sources were used (on-line CVs, university web sites, etc.).

The remaining 20% of cases are largely traceable to pre-existing errors in the bibliometric source, in particular in four instances the address listed in the WoS is different from that indicated in the title area of the actual publication.

The data in Table 5 thus reveal that 60% of the identified false negatives are due to errors either by the authors or present in the external database, and do not result from the algorithm. Net of such errors, the number of false negatives (24) is actually less than for false positives (28), and the true recall (96.2%) is greater than precision (95.6%).

[Table 5]

**Test based on the University of Milan Institutional Research Archives**

The second type of test was conducted by comparison with data available from the University of Milan's Institutional Research Archives (AIR)[23]. This university is the fourth largest in Italy, with over 2,500 researchers, of which 70% belong to the hard science SDSs. The examination of this case is also interesting because Milan is also the seat of four other universities, which leads to a higher probability for presence of errors in the controlled vocabulary used in the address filter.

The AIR is a database in which the officially recognized researchers at the University of Milan enter data on their personal scientific product. The university's library service checks the data inserted, corrects them if necessary, and further integrate them with results from regular querying of the WoS. The AIR database is sufficiently complete for the years 2005-2007: for this period it registers 13,428 authorships from the WoS source. However, an in-depth preliminary analysis revealed a series of records with errors[24]:
- 271 duplicate authorships;
- 674 authorships erroneously indicated as being sourced from the WoS;
- 411 authorships involving individuals who did not hold a faculty role as of

---

[22] Such percentages are purely indicative and do not have any statistical value with respect to the population.

[23] http://air.unimi.it/

[24] The count of errors in AIR is beyond the scope of the present work, however it can suggest a general idea of the level of accuracy of databases compiled "from the bottom up". Further, it can present a useful reference with respect to the precision and recall of the proposed algorithm, which is instead based on a top-down strategy.

December 30 of the year preceding the publication date.

To this can be added:
- 599 authorships concerning publications which were not an "article" or a "review",
- 301 authorships concerning publications with more than 50 authors[25].

In the end, for the purposes of this comparison, the AIR database presents 11,172 valid authorships. In comparison, the proposed algorithm applied directly to the WoS identifies 10,724 authorships, or 96%. This number can not be safely assumed as a true measure of recall since among the disambiguated identities, other than false negatives, there could also be potential false positives.

In this regard, the test proceeded to a further activity of manual checking of a series of uncertain cases. In particular, to identify possible false negatives, a check was made of the scientific product identified from the WoS for:
- Researchers who were non-productive according to the proposed algorithm but had at least one publication in the AIR;
- Researchers with a difference of greater than two between the publications assigned by the proposed algorithm and those listed in the AIR, or in any case with a difference of greater than 30% from the AIR data.

For the identification of false positives, a manual check of scientific product from the WoS source was conducted for:
- Researchers who were non-productive according to the AIR but had two or more publications attributed by the algorithm;
- Researchers with a difference of greater than two between the publications assigned by the proposed algorithm and those listed in the AIR, or in any case with a difference of greater than 30% from the AIR data.

Table 6 presents summary data of the cases selected for manual checking and the errors identified.

[Table 6]

It can be seen that, in number identified, the false negatives exceed the false positives, potentially demonstrating that the performance for precision is slightly superior to that for recall. Assuming the status of true positives for all 11,712 AIR records used for the comparison[26], the measures of precision and recall result as respectively 96.4% and 94.3%, with f-measure of 95.3%[27].

These values are fully comparable with those obtained in the previous section, based on a sample observation of the entire national dataset.

**Conclusions**

Many nations are establishing a regular practice of conducting periodic national exercises for evaluation of research. These exercises are for the most part based on the peer review approach, which appears to suffer from serious limitations, but which has still been preferred over other alternatives. For the hard sciences, where scientific publication is the principle form for codifying the diffusion of research results, the bibliometric approach appears to offer definite advantages. The bibliometric approach is reliable (based on objective data, which are quantitative and homogenous, concerning the entire scientific product of an organization), offers rapid and economical implementation, and can be not at all invasive. Further, census and evaluation on the basis of the scientific product of single identifiable individuals appears fundamental to the effectiveness of the evaluation system. Such census permits measurement of the research productivity at all levels desired: single individuals, but also departments, scientific sectors, disciplinary areas, faculties and entire organizations. Further, measurement of productivity of single authors is useful not only in the realm of government-mandated exercises for evaluation on a national scale, but also at the level of single organizations, for selective funding allocation.

---

[25] Such publications are considered anomalous and are excluded a priori from the disambiguation algorithm.
[26] Actually, there could still be residual false positives and negatives included among these records.
[27] As previously, these values do not indicate a "true" recall for the algorithm, since many of the false negatives counted result from errors in the external source, or from other causes extraneous to the logic of the algorithm.

However, technical limitations inherent in the bibliometric databases currently available have held back the use and diffusion of bibliometrics as a non-invasive support system for the evaluation of research. These limitations are primarily traced to the difficulties involved in correctly identifying the true authors of each publication, particularly because of homonyms among names and variations in the way individual authors indicate their name and affiliation.

The current work has concentrated on this problem and has proposed a possible solution that foresees the integration of structured data with a bibliometric database, through a procedure that takes a very natural course. In the present case we have used an external source of reference information that provides data on the affiliation and research area of each Italian academic scientist to disambiguate the true identity of the authors of publications listed in the Thomson Reuters I-NCR. The external database, albeit crucial for the applicability of our algorithm, is not a particularly critical resource. National research systems are typically composed of communities that can be easily identified, and gathering data to build a comprehensive external database should not require a significant human effort.

Compared to similar attempts reported in the literature, and especially to state-of-the-art unsupervised approaches, the proposed algorithm presents definite advantages in terms of scalability and expandability, and above all it offers practical and very easy application. Finally, the performance for precision and recall seems completely adequate for the typical needs of bibliometric applications, and still offer further scope for improvement. The development and application of the algorithm described has allowed the authors to set up a unique national bibliometric database, apt to support large scale research evaluations at the level of the single scientist. The hope is that other countries as well set up similar databases to make international comparisons possible.

**References**

Abramo, G., D'Angelo, C.A. and Viel, F. (2010). Peer review research assessment: a sensitivity analysis of performance rankings to the share of research product evaluated, *Scientometrics*, DOI: 10.1007/s11192-010-0238-0.
Abramo, G., D'Angelo, C.A., Caprasecca, A. (2009). Allocative efficiency in public research funding: can bibliometrics help? *Research Policy*, 38(1), 206-215.
Abramo, G., D'Angelo, C.A., Di Costa, F. (2008a). Assessment of sectoral aggregation

distortion in research productivity measurements. *Research Evaluation*, 17(2), 111-121.

Abramo, G., D'Angelo, C.A., Pugini, F. (2008b). The measurement of Italian universities' research productivity by a non parametric-bibliometric methodology. *Scientometrics*, 76(2), 225-244.

Aksnes, D.W. (2008). When different persons have an identical author name. How frequent are homonyms? *Journal of the American Society for Information Science and Technology*, 59(5), 838-841.

Balconi, M., Breschi, S., & Lissoni, F. (2004). Networks of inventors and the role of academia: an exploration of Italian patent data. *Research Policy*, 33(1), 127-145.

Cornell, L.L. (1982). Duplication of Japanese names: a problem in citations and bibliographies. *Journal of the American Society for Information Science and Technology*, 33(2), 102-104.

Culotta, A., Kanani, P., Hall, R., Wick, M., & McCallum, A. (2007). Author disambiguation using error-driven machine learning with a ranking loss function. *Proceedings of the 6th International Workshop on Information Integration on the Web (IIWeb 2007)* (pp. 32-37). Menlo Park, California: AAAI Press.

Han, H., Giles, L., Zha, H., Li, C., & Tsioutsiouliklis, K. (2004). Two supervised learning approaches for name disambiguation in author citations. *Proceedings of the 4th ACM/IEEE-CS Joint Conference on Digital Libraries (JCDL 2004)* (pp. 296-305). New York, NY, USA: ACM.

Han, H., Xu, W., Zha, H., & Giles, C. L. (2005a). A hierarchical naive Bayes mixture model for name disambiguation in author citations. *Proceedings of the 20th Annual ACM Symposium on Applied Computing (SAC 2005)* (pp. 1065-1069). New York, NY, USA: ACM.

Han, H., Zha, H., & Giles, C. L. (2005b). Name disambiguation in author citations using a K-way spectral clustering method. *Proceedings of the 5th ACM/IEEE-CS Joint Conference on Digital Libraries (JCDL 2005)* (pp. 334-343). New York, NY, USA: ACM.

Harman, G. (2000). Allocating research infrastructure grants in post-binary higher education systems: British and Australian approaches. *Journal of Higher Education Policy and Management*, 22(2), 11-126.

Huang, J., Ertekin, S., & Giles, C. (2006). Efficient Name Disambiguation for Large-Scale Databases. *Proceedings of the 10th European Conference on Principles and Practice of Knowledge Discovery in Databases (PKDD 2006)* (pp. 536-544). Berlin, Germany: Springer.

Iversen, E. J., Gulbrandsen, M., & Klitkou, A. (2007). A baseline for the impact of academic patenting legislation in Norway. *Scientometrics*, 70(2), 393-414.

Kanani, P., McCallum, A., & Pal, C. (2007). Improving author coreference by resource-bounded information gathering from the web. *Proceedings of the 20th International Joint Conference on Artificial Intelligence (IJCAI 2007)* (pp. 429-434). Menlo Park, California: AAAI Press.

Li, X., Morie, P., & Roth, D. (2005). Semantic integration in text: from ambiguous names to identifiable entities. *AI Magazine*, 26(1), 45-58.

McRae-Spencer, D. M., & Shadbolt, N. R. (2006). Also by the same author: AKTiveAuthor, a citation graph approach to name disambiguation. *Proceedings of the 6th ACM/IEEE-CS Joint Conference on Digital Libraries (JCDL 2006)* (pp. 53-54). New York, NY, USA: ACM.

Meyer, M. (2003). Academic patents as an indicator of useful research? A new approach to measure academic inventiveness. *Research evaluation*, 12(1), 17-27.

On, B., Lee, D., Kang, J., & Mitra, P. (2005). Comparative study of name disambiguation problem using a scalable blocking-based framework. *Proceedings of the 5th ACM/IEEE-CS Joint Conference on Digital Libraries (JCDL 2005)* (pp. 344-353). New York, NY, USA: ACM.

Radicchi, F., Fortunato, S., Markines, B., Vespignani, A. (2009). Diffusion of scientific credits and the ranking of scientists. *Physical Review E*, 80, 056103.

Soler, J. (2007). Separating the articles of authors with the same name. *Scientometrics*, 72(2), 281-290.

Song, Y., Huang, J., Councill, I. G., Li, J., & Giles, C. L. (2007). Efficient topic-based unsupervised name disambiguation. *Proceedings of the 7th ACM/IEEE-CS Joint Conference on Digital Libraries (JCDL 2007)* (pp. 342-351). New York, NY, USA: ACM.

Tan, Y. F., Kan, M. Y., & Lee, D. (2006). Search engine driven author disambiguation. *Proceedings of the 6th ACM/IEEE-CS Joint Conference on Digital Libraries (JCDL 2006)* (pp. 314-315). New York, NY, USA: ACM.

Torvik, V. I., Weeber, M., Swanson, D. R., & Smalheiser, N. R. (2005). A probabilistic similarity metric for Medline records: A model for author name disambiguation. *Journal of the American Society for Information Science and Technology*, 56(2), 140-158.

Van Raan, A. F. J. (2008). Scaling rules in the science system: Influence of field-specific citation characteristics on the impact of research groups. *Journal of the American Society for Information Science and Technology*, 59(4), 565-576.

Wooding, S., Wilcox-Jay, K., Lewison, G., & Grant, J. (2006). Co-author inclusion: A novel recursive algorithmic method for dealing with homonyms in bibliometric analysis. *Scientometrics*, 66(1), 11-21.

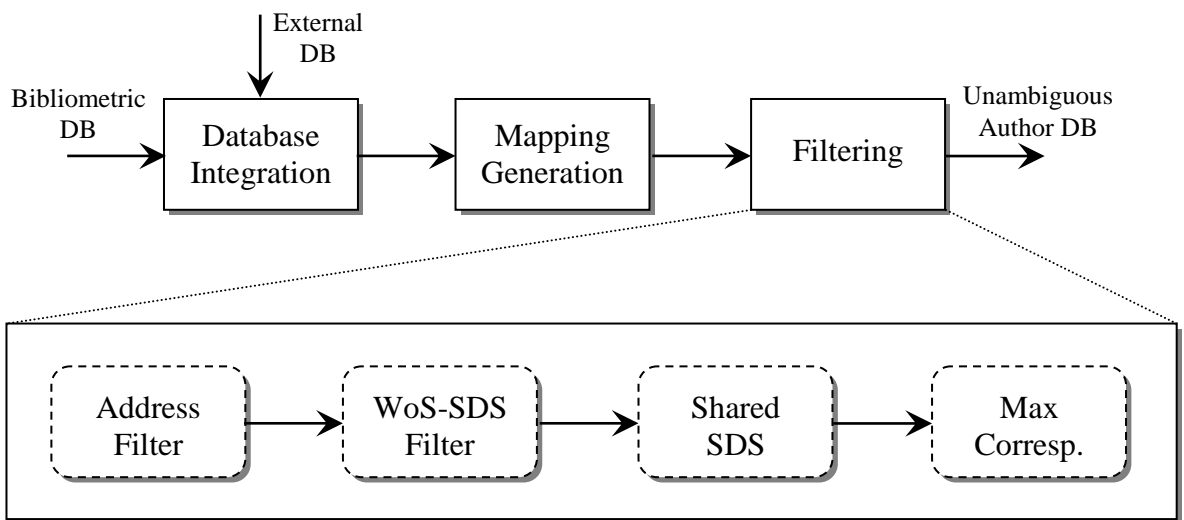

**Figure 1: Flowchart of the proposed algorithm**

|  | Macro-categories | Publications |
|---|---|---|
| Hard sciences | Mathematics | 10,804 |
|  | Physics | 45,292 |
|  | Chemistry | 21,842 |
|  | Earth and Space Sciences | 12,894 |
|  | Biology | 34,334 |
|  | Biomedical Research | 34,679 |
|  | Clinical Medicine | 61,827 |
|  | Engineering | 39,738 |
|  | *Sub total* | *261,410 (96.4%)* |
| Social Sciences | Economics | 2,369 |
|  | Law, political and social sciences | 1,752 |
|  | Psychology | 2,133 |
|  | *Sub-total* | *6,254 (2.3%)* |
| Art & Humanities | Art & Humanities | 2,221 (0.8%) |
|  | Multidisciplinary Sciences | 1,411 (0.5%) |
|  | *Total* | *271,296* |

*Table 1: Publications in the I-NCR by macro disciplinary area (totals 2001-2007)*

|  | UDA | Total Staff |
|---|---|---|
| Hard Sciences | Mathematics and information science | 3,413 |
|  | Physics | 2,617 |
|  | Chemistry | 3,335 |
|  | Earth sciences | 1,290 |
|  | Biological sciences | 5,381 |
|  | Medical sciences | 11,480 |
|  | Agriculture and veterinary sciences | 3,303 |
|  | Civil engineering and architecture | 3,930 |
|  | Industrial and information engineering | 5,105 |
|  | Subtotal | 39,854 (63.7%) |
| Social Sciences | Economics and statistics | 4,679 |
|  | Law and legal sciences | 5,017 |
|  | Social and political sciences | 1,709 |
|  | Subtotal | 11,405 (18.2%) |
| Art & Humanities | Classics, literature and languages, art history | 6,041 |
|  | History, philosophy, psychology, education | 5,259 |
|  | Subtotal | 11,300 (18.1%) |
|  | Total | 62,559 |

*Table 2: Research staff in the Italian academic system (as of 31/12/2006)*

| Step | Papers | Identities | Identities per paper |
|---|---|---|---|
| Mapping generation | 233,661 | 1,339,024 | 5.731 |
| Address filter | 191,775 (-17.9%) | 461,003 (-65.6%) | 2.404 (-58.1%) |
| WoS-SDS filter | 188,934 (-1.5%) | 421,387 (-8.6%) | 2.230 (-7.2%) |
| Shared SDS filter | 188,934 (0%) | 412,881 (-2.0%) | 2.185 (-2.0%) |
| Max WoS-SDS correspondence filter | 188,934 (0%) | 406,534 (-1.5%) | 2.152 (-1.5%) |

*Table 3: Identities per paper through the various steps of the algorithm (bibliometric dataset extracted from the Italian university publications indexed in the WoS 2001-2007)*

| Population (total authorship disambiguated) | N | 406,534 |
|---|---|---|
| Confidence level | Z | 2.33 (98%) |
| Sampling error | e | 3% |
| Population heterogeneity | p | 12% |

*Table 4: Sampling data*

| Causes of false negative generation | Frequency |
|---|---|
| Error by the author in indicating his or her address | 16 (40%) |
| Error from WoS-SDS filtering | 12 (30%) |
| Error in address recognition | 4 (10%) |
| Error in WoS address listing | 4 (10%) |
| Error in WoS listing of author name | 3 (7.5%) |
| Error in matching WoS name-CINECA name | 1 (2.5%) |
| Total | 40 |

*Table 5: Frequency of causes for generation of false negatives*

| Type of check | Researchers checked | Pairs checked | Errors identified |
|---|---|---|---|
| False negatives | 258 | 1,508 | 634 |
| False positives | 203 | 2,001 | 391 |

*Table 6: Errors identified in the disambiguated pairs by means of comparing against the University of Milan Institutional Research Archives*